%% file: skeleton_arXiv.tex
\documentclass{PoS}

\usepackage{booktabs} 
\usepackage{lineno} 

\title{Rare decays of B-hadrons}

\ShortTitle{Rare decays of B-hadrons}

\author{\speaker{Carla Marin Benito}\thanks{On behalf of the LHCb collaboration}\\
        Laboratoire de l'Acc\'el\'erateur Lin\'eaire \\
        E-mail: \email{carla.marin@cern.ch}}

\abstract{Rare decays of \bquark-hadrons provide high sensitivity to New Physics effects. Several deviations with respect to the Standard Model predictions have been 
	observed in recent years, leading to significant tensions in global fit analyses. It is thus crucial to update the existing measurements 
	and study new decay modes to confirm the pattern. The latest results from \lhcb and Belle on radiative, semileptonic penguin, lepton universality and lepton flavour violation decays are presented.}

\FullConference{XXIX International Symposium on Lepton Photon Interactions at High Energies - LeptonPhoton2019\\
		August 5-10, 2019\\
		Toronto, Canada}

\def\PLB{{\em Phys. Lett.}  B}

\def\PRD{{\em Phys. Rev.} D}

\def\EPJC{{\em Eur. Phys. J.} C}

\input{symbols}

\begin{document}

\section{Introduction}

\begin{figure}
	\centering
	{\includegraphics[width=0.4\linewidth]{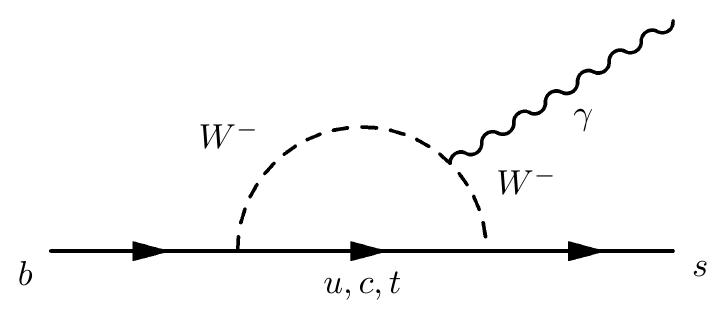}}
	\caption[]{Feynman diagram of the FCNC \bsg transition in the SM.}
	\label{fig:btosg_diag}
\end{figure}

Rare decays of b-hadrons are Flavor-Changing Neutral-Currents (FCNC), which are forbidden 
at tree level in the Standard Model (SM) and are thus very suppressed.  As such, they are 
very sensitive to potential new particles entering the loops virtually and affecting properties of the decays 
such as branching fractions and angular distributions. Consequently, the measurement of these processes allows to probe higher scales 
than direct searches. As an example, the Feynman diagram of the FCNC \bsg transition in 
the SM is shown in Fig.~\ref{fig:btosg_diag}.

From the theoretical viewpoint, these decays can be described in a model-independent way by means 
of the Operator Product Expansion. The effective Hamiltonian of the theory is expressed as a sum 
over all the possible operators, which are modulated by a set of coefficients, known as Wilson coefficients ($C_i$). The most relevant effective couplings are those parametrising the vector ($C_9$),
vector-axial ($C_{10}$) and photon mediated ($C_7$) operators.
The Wilson coefficients can be computed in the SM and compared to the values extracted from global fits to experimental data. 
Any significant deviation from the SM calculations taking into account both experimental and theory 
uncertainties would point to New Physics (NP) effects.

In recent years several deviations with respect to SM predictions have been observed in this type of processes. 
Differential branching fraction (\BR) measurements in $b \to s \ell \ell$ decays exhibit a trend towards values lower than the SM predictions in the di-lepton mass squared, $q^2$, 
region below the charmonium threshold~\cite{LHCb-PAPER-2014-006,LHCb-PAPER-2015-023,LHCb-PAPER-2015-009},
although theoretical predictions for these observables are affected by large hadronic uncertainties. 
Deviations have been also observed in the theoretically cleaner $P'_5$ angular observable in the 
\BdToKstmm decay~\cite{LHCb-PAPER-2015-051}. 
Lepton Flavor Universality tests, which are complementary to these measurements and 
provide theoretically very precise observables, also show a deviation with respect to the universal 
SM prediction~\cite{LHCb-PAPER-2014-024,LHCb-PAPER-2017-013}. 

In view of all these anomalies in rare b-hadron decays, it is crucial to update the previous measurements with more data
and study new complementary modes. The most recent results from LHCb and Belle on rare \bquark-hadron decays 
are reported in the following.

\section{Radiative decays}
Radiative decays of \bquark-hadrons are mediated by the \bsg quark-level transition, governed by the $C^{(')}_7$ Wilson coefficient. 
The $C_7$ coefficient is very constrained already by the measurement of the branching fraction of the inclusive $\decay{\B}{X_{s}\g}$ process 
and of direct CP asymmetry in the \BdKstGam decay. However, the presence of right-handed currents beyond the standard model in this transition, 
entering in the $C'_7$ coefficient, has only been mildly constrained so far. 
Experimentally, this can be tested through a measurement of the 
polarisation of the emitted photon, which is predicted to be left-handed in the SM, up to corrections of the order $m_{s}/m_{b}$.
Current studies focus on the measurement of observables sensitive to the photon polarisation to improve the constraints on potential new physics
entering $C'_7$ . 

\subsection{Photon polarisation in \bsphig}
\begin{figure}
	\centering
	{\includegraphics[width=0.32\linewidth]{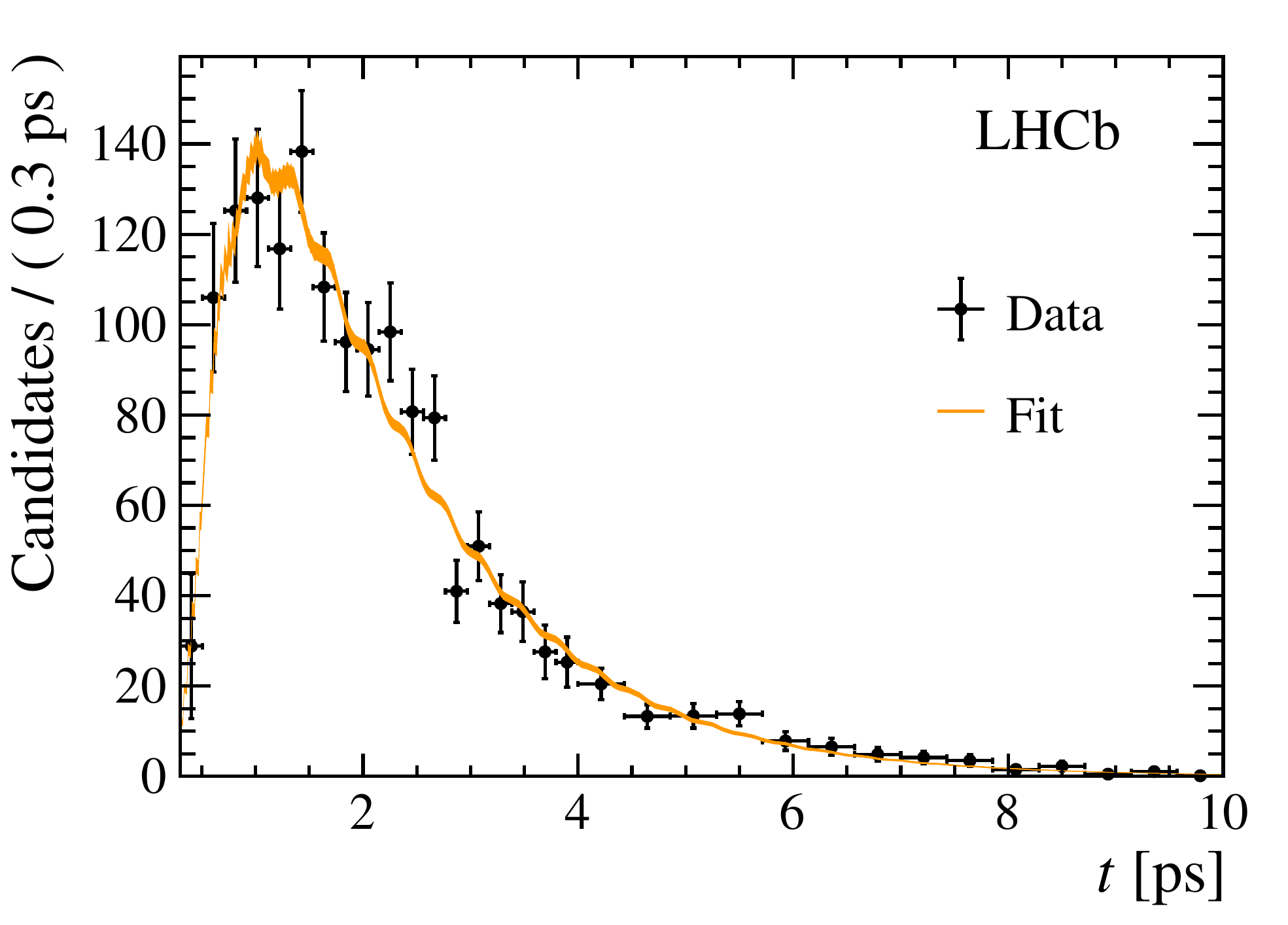}}
	{\includegraphics[width=0.32\linewidth]{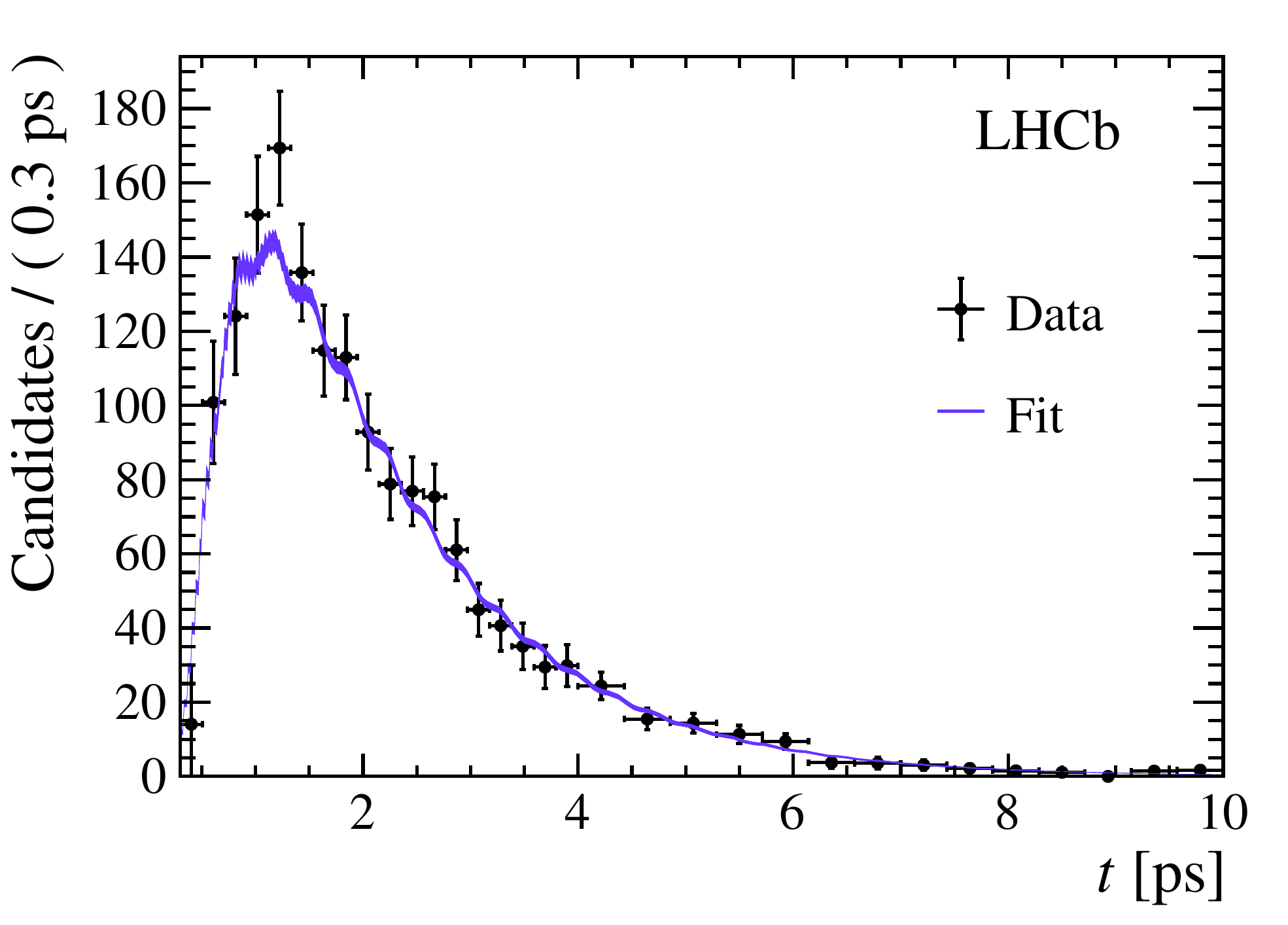}}
	{\includegraphics[width=0.32\linewidth]{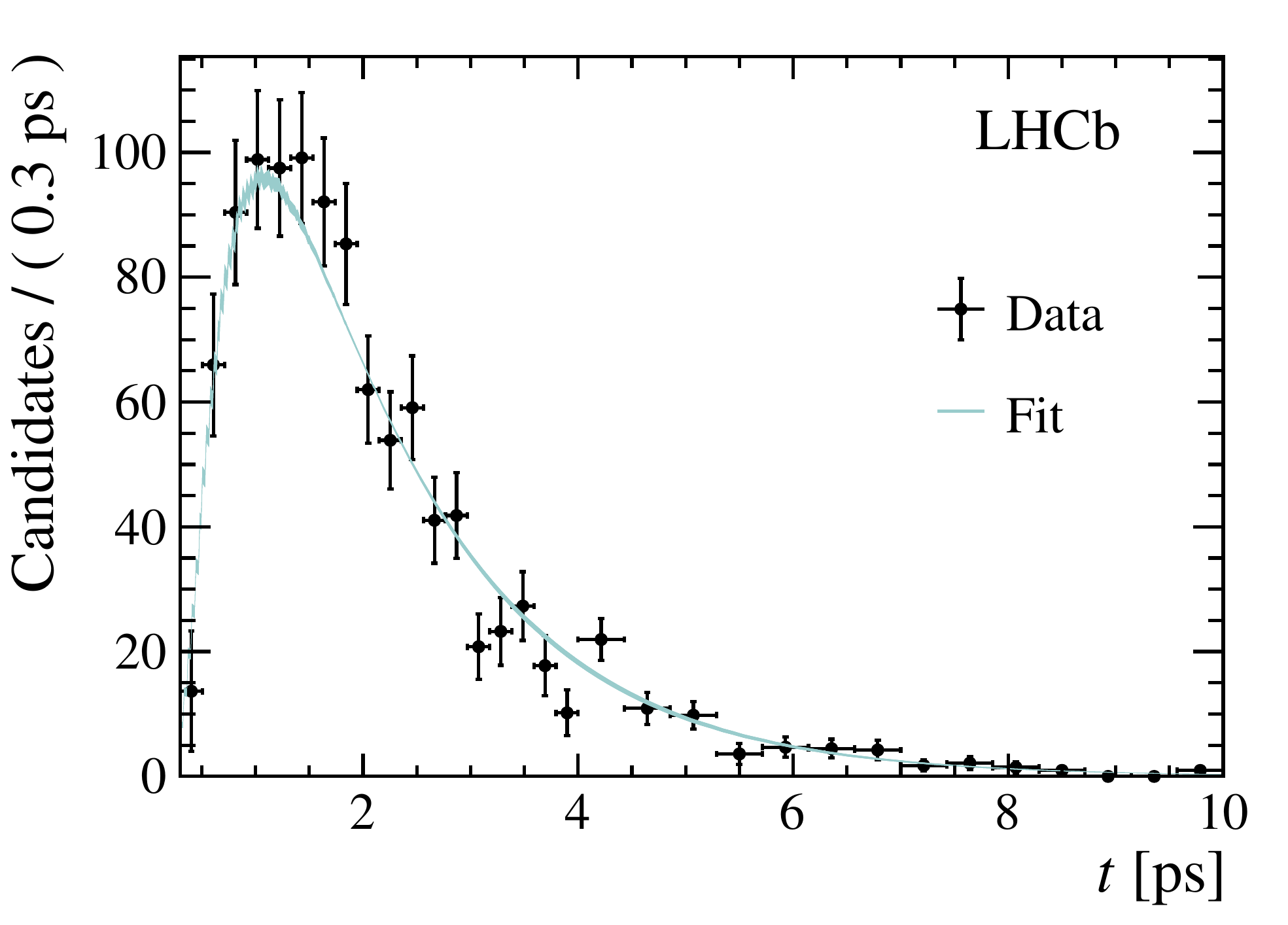}}
	\caption[]{Time-dependent decay rate of \bsphig candidates for events tagged as \Bs (left), 
		$\overline\Bs$ (middle) and with no tagging information (right). The black points show the data and the results 
	of the fit are represented by the solid curves.}
	\label{fig:bsphig}
\end{figure}

The time-dependent rate of a \Bs meson decay to a CP even final state is given by the expression:
\begin{equation}
\label{eq:bsphig_th}
	\Gamma(t) \propto \exp^{-\Gamma_{s}t} \left[ \cosh \left( \frac{\Delta\Gamma_{s}t}{2} \right) - {A}^{\Delta} \sinh \left(\frac{\Delta\Gamma_{s}t}{2} \right) \pm
	{C}_{CP} \cos{\left( \Delta m_{s}t \right)} \mp {S}_{CP} \sin{\left( \Delta m_{s}t \right)} \right]
\end{equation}
where $\Delta\Gamma_{s}$ and $\Delta m_{s}$ are the width and mass differences between the 
two mass eigenstates and $\Gamma_{s}$ is their average width. The last two terms change 
the sign of their contribution depending on whether the produced particle is a \Bs meson or an 
anti-\Bs meson and thus one needs to know the flavour of the hadron at production to measure them.
The coefficients ${A}^{\Delta}$ and ${S}_{CP}$ are sensitive to the photon helicity 
and weak phases, while ${C}_{CP}$ is related to CP violation in the decay. 
Exploiting the decay \bsphig, LHCb measured ${A}^{\Delta} = -0.98 ^{+0.46+0.23}_{-0.52-0.20}$ using a dataset corresponding to 
$3\invfb$ collected in Run 1 of the LHC, in an analysis where no information 
on the flavour of the initial \Bs was used~\cite{LHCb-PAPER-2016-034}. 
Earlier this year, an updated measurement including flavour tagging information was published~\cite{LHCb-PAPER-2019-015}. The time-dependent decay rate of events tagged as \Bs,
$\overline\Bs$ or with no tagging information is fit separately to extract the coefficients entering Eq.~\ref{eq:bsphig_th}, 
as shown in Fig.~\ref{fig:bsphig}. The obtained results are: 
\begin{eqnarray}
\label{eq:bsphig_res}
	{A}^{\Delta} &=& -0.67 ^{+0.37}_{-0.41} \pm 0.17 \nonumber\\
	{C}_{CP} &=& 0.11 \pm 0.29 \pm 0.11 \nonumber\\
	{S}_{CP} &=& 0.43 \pm 0.30 \pm 0.11,
\end{eqnarray}
where the first uncertainty is statistical and the second systematic. The results are compatible with 
the absence of right-handed currents and CP violation. This is the first measurement of the 
${C}_{CP} $ and ${S}_{CP} $ coefficients in the \Bs system. 

\subsection{First observation of \LbLzGam}
The \LbLzGam decay is also a FCNC \bsg transition, which was previously unobserved. The SM prediction for its branching ratio stands in the range 
$10^{-7}-10^{-5}$, with the large uncertainty originating from the computation of form factors~\cite{Wang:2008sm,Mannel:2011xg,Gan:2012tt,Faustov:2017wbh}. The best limit on this observable 
was set by CDF at $\BRlblg < 1.9 \times 10^{-3}$ at $90\%$ CL~\cite{Acosta:2002fh}, which leaves large room for experimental improvement to reach the SM prediction.
This decay offers direct access to the photon polarisation 
in \bsg decays through the angular distribution of the final state particles and can thus probe the existence of right-handed currents~\cite{Mannel:1997xy}.
LHCb has performed a search for this decay mode using the dataset collected in 2016, which corresponds to $1.7 \invfb$ of integrated luminosity~\cite{LHCb-PAPER-2019-010}.

The particular topology of this decay poses a great challenge for its reconstruction at LHCb. The \Lb decay vertex, typically exploited to suppress
prompt background, cannot be reconstructed in this case due to the long lifetime of the \Lz baryon and the lack of information on the 
direction of the photon detected in the electromagnetic calorimeter. Consequently, a dedicated online and offline reconstruction was 
developed to be able to study this mode, where the \Lb momentum is computed from the direct sum of the \Lz and \g momenta, without a vertex fit. 
A large combinatorial background is expected due to the impossibility of applying tight requirements on the \Lb decay vertex
and is mitigated with a high performance BDT trained using the XGBoost~\cite{Chen:2016:XST:2939672.2939785} algorithm. 
Neutral particle identification tools~\cite{CalvoGomez:2042173} are exploited to reject potential background from \piz misidentification.
Other sources of background, apart from a small contamination from \LbLzEta decays with \decay{\eta}{\g \g}, are found to be negligible.

The well-known \BdKstGam decay is used as normalisation mode to extract a branching ratio measurement 
using the recent measurement of the \Lb hadronisation fraction ratio at $13\tev$ by \lhcb ~\cite{LHCb-PAPER-2018-050}. 
The normalisation mode and intermediate state branching fractions are taken from the PDG~\cite{PDG2018} and the efficiencies, $\epsilon$, are computed from simulation and calibration samples. 

\begin{figure}
	\centering
	{\includegraphics[width=0.32\linewidth]{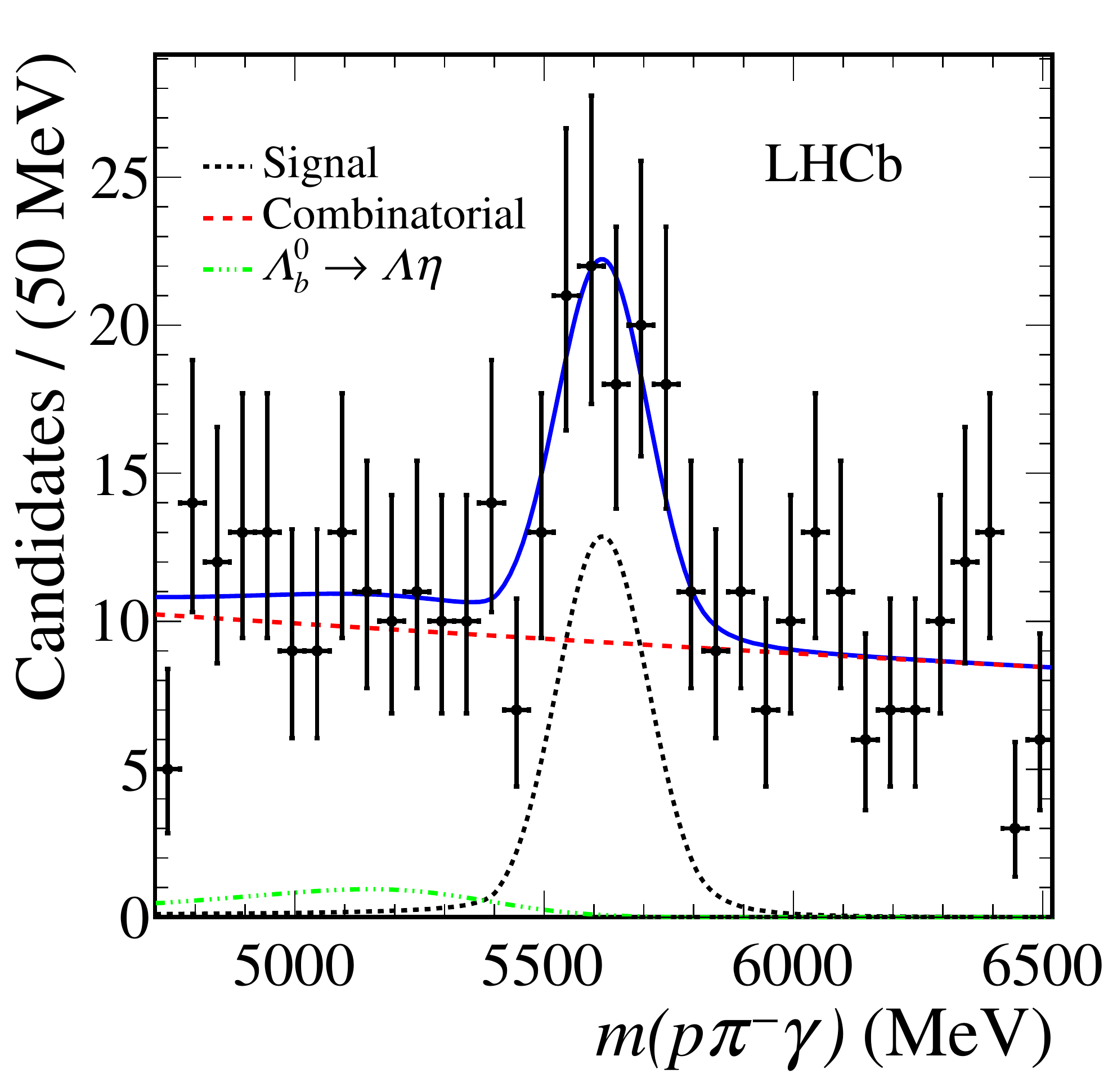}}
	{\includegraphics[width=0.32\linewidth]{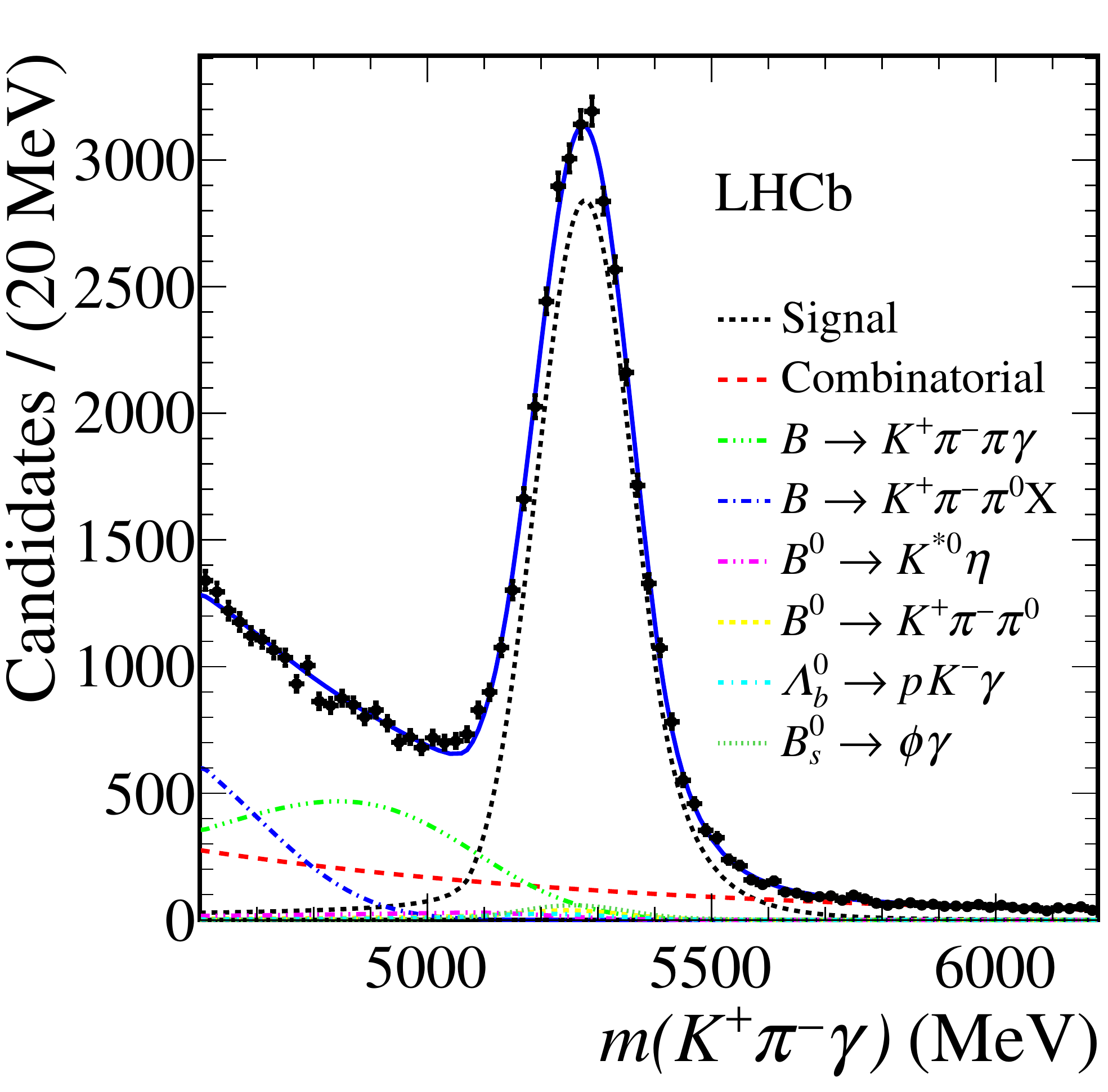}}
	\caption[]{Invariant mass distribution of selected \LbLzGam (left) and \BdKstGam (right) candidates (black dots). 
		The result of a simultaneous unbinned maximum likelihood fit is shown by the blue curve, with different contributions 
		represented by different style and color curves, as described in the legend.}
	\label{fig:LbLzGam}
\end{figure}

The results of a simultaneous unbinned maximum likelihood fit to signal and normalisation candidates are shown in Fig.~\ref{fig:LbLzGam}. 
A clear signal peak of $65 \pm 13$ events is observed, together with $32670 \pm 290$ \BdKstGam events. 
The significance of the signal excess is found to be $5.6 \sigma$, which represents the first observation of the \LbLzGam decay mode. 
The branching fraction is measured to be
\begin{equation}
\BRLbLzGam = (7.1 \pm 1.5 \pm 0.6 \pm 0.7) \times 10^{-6},
\end{equation}
where the first uncertainty is statistical, the second systematic and the third arises from external inputs, which are dominated by the systematic uncertainties on the 
determination of the ratio of hadronisation fractions. 
The measurement is well within the range of SM predictions and can be used to determine the 
\LbtoLz form factors at the photon pole. With a larger dataset this decay mode can be exploited to obtain a measurement of the photon polarisation in radiative b-baryon decays.

\section{Semileptonic penguin decays}

Semileptonic \bsll transitions are mediated by the $C_7$, $C_9$ and $C_{10}$ Wilson coefficients in the SM, 
in different proportions depending on the $q^2$ region. The di-lepton spectrum is dominated 
by the narrow charm resonances \jpsi and \psitwos that are typically excluded in analyses and used as control 
modes instead. The main focus in this sector is on angular analyses and lepton flavour universality tests, 
both on the update of previous analyses with more data and improved experimental techniques and 
on the measurement of these observables in new decay modes not explored before. 

\subsection{Angular analysis of \LbToLzmm}

\begin{figure}
	\centering
	{\includegraphics[width=0.6\linewidth]{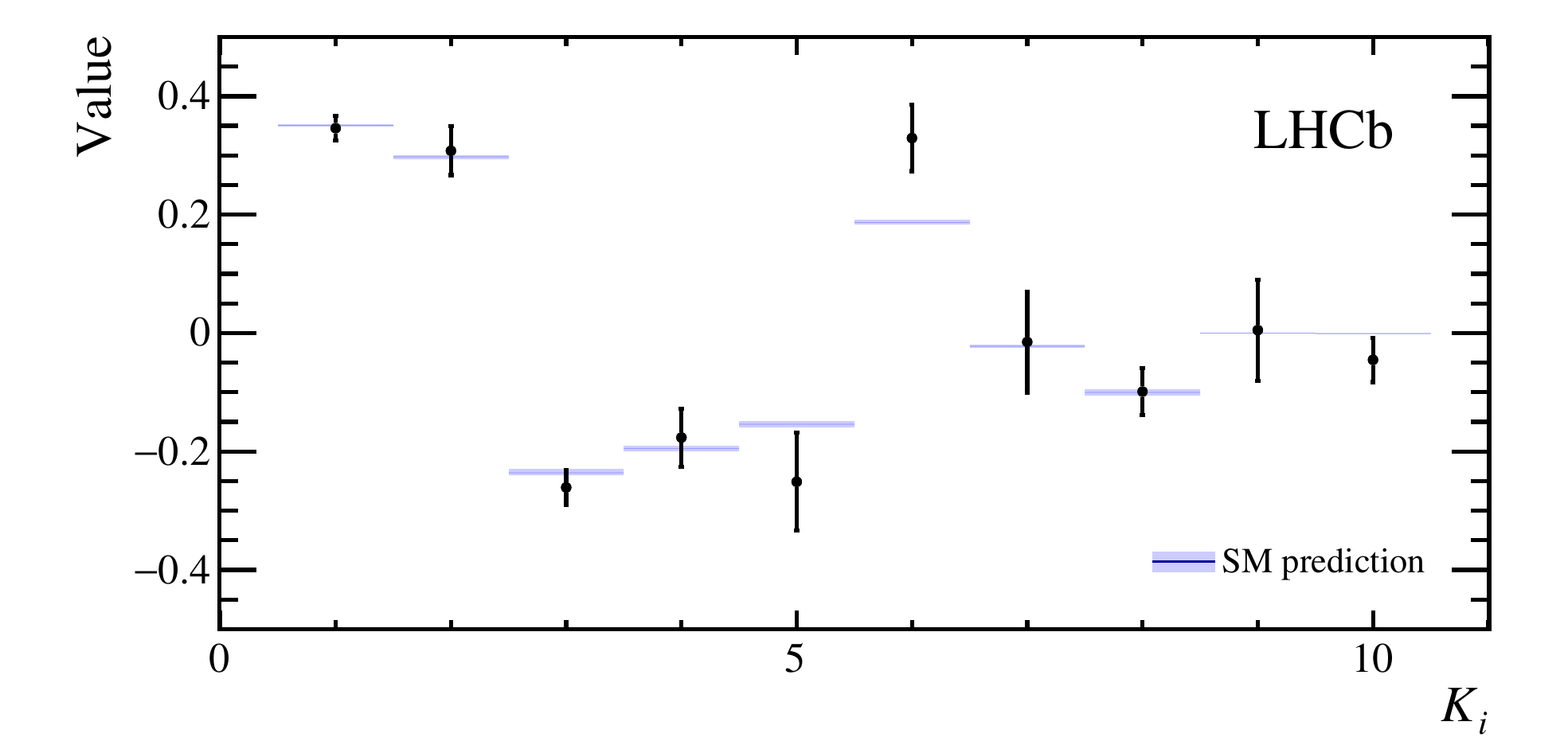}}
	\caption[]{Angular coefficients of the \LbToLzmm decay obtained from the fit to data (black dots) and SM prediction (blue bands).}
	\label{fig:LbLzmm}
\end{figure}

The \LbToLzmm decay is a \bsll transition complementary to \BdToKstmm with a richer angular distribution due to the half-integer 
spin of the baryons and the potential initial polarisation of the \Lb hadron. The analysis~\cite{LHCb-PAPER-2018-029} is performed using the data recorded by LHCb in 
the period 2011 -- 2016, corresponding to an integrated luminosity of $5 \invfb$, and focuses on the low recoil region, 
$15 < q^{2}  <20 \gev^2/c^4$, where most of the signal lies. The angular analysis exploits the method of moments~\cite{Blake:2017une} 
to measure for the first time the full set of angular observables for this decay.
The 5-dimensional angular distributions are described by the sum of 35 angular functions, 
\begin{equation}
\frac{d^5\Gamma}{d\Omega} = \frac{3}{32\pi^2}\sum_{i}^{34}K_i f_i(\Omega),
\end{equation}
where $K_i$ are the angular coefficients extracted from the fit to data. The results for these parameters are shown in Fig.~\ref{fig:LbLzmm}. 
All the coefficients are compatible with the SM predictions computed with the \texttt{EOS}~\cite{EOS} software 
using the \Lb production polarisation measured by LHCb~\cite{LHCb-PAPER-2012-057}. 
The $K_{11,...,34}$ coefficients are found to be compatible with zero, i.e., with the absence of initial \Lb polarisation.
Forward-backward asymmetries in the lepton and hadron systems and between the two systems can be derived from the combination of different 
coefficients. They are found to be
\begin{eqnarray*}
	A_{FB}^{\ell} &=& \frac{3}{2}K_{3} = -0.39\pm0.04\pm0.01 \nonumber\\
	A_{FB}^{h} &=& K_{4} + \frac{1}{2}K_{5} = -0.30\pm0.05\pm0.02 \nonumber\\
	A_{FB}^{h\ell} &=& \frac{3}{4}K_{6} = +0.25\pm0.04\pm0.01,\\
\end{eqnarray*}
which are also in agreement with the SM predictions. 

\subsection{Lepton Universality Tests}
Lepton universality tests in \bsll transitions involve the comparison of muon and electron final states that share the same hadronic content, 
which in the SM are predicted to behave exactly equal, except for the lepton masses. Experimentally, electrons and muons are detected very differently, 
specially in a high momentum and occupancy environment such as the LHC, causing reconstruction and selection differences 
than need to be carefully controlled. 

At LHCb, the main differences arise from the way these events are triggered and from the Bremsstrahlung losses that 
electrons suffer as they traverse the detector material. At the hardware trigger level, electrons are selected as high 
energetic clusters in the electromagnetic calorimeter (ECAL), while muons are distinguished by the traces they leave in the 
muon stations. Due to the different occupancies of both detectors, a larger transverse momentum threshold is needed 
in the ECAL to control the rate of selected events. This is mitigated by exploiting other parts of the event to trigger on. 
In particular, the hadron from the \bsll decay of interest or any high energetic signature from the other \bquark produced 
in the event are used.

A procedure is in place to correct the electron track momentum from potential Bremsstrahlung loses. 
The direction of the track before the magnet is propagated to the ECAL surface and the momenta of photon clusters 
near the extrapolated track position are added to the one measured by the tracking. This provides a good recovery 
of a large fraction of the emitted energy but some signal photons can be missed while others originating from 
different processes can be incorrectly added, causing a degradation of the mass resolution for electron modes 
in comparison to muon ones.

The ratio of branching fractions of a given \bsll process with electrons or muons in the final state is predicted in the SM to be 
\begin{equation}
	R_{H} = \frac{\BR(\decay{H_b}{H_s \mumu})}{\BR(\decay{H_b}{H_s \epem})} = 1
\end{equation}
with a $1\%$ accuracy~\cite{bordone}. Experimentally, one measures the yields of these two decays, which can be 
related to the ratio of branching fractions through the reconstruction and selection efficiencies:
\begin{equation}
R_{H} = \frac{N(\decay{H_b}{H_s \mumu})}{N(\decay{H_b}{H_s \epem})} \times 
\frac{\epsilon(\decay{H_b}{H_s \epem})}{\epsilon(\decay{H_b}{H_s \mumu})}.
\end{equation}
Event yields are obtained from invariant mass fits to data candidates, while efficiencies are computed from 
simulation and calibration samples. In order to validate the correct description of the data provided by the 
simulation, the corresponding \jpsi modes, which have been tested accurately in the past to be flavour universal~\cite{PDG2018}, 
are exploited. The ratio
\begin{equation}
r_{\jpsi} = \frac{\BR(\decay{H_b}{H_s \jpsi(\mumu)})}{\BR(\decay{H_b}{H_s \jpsi(\epem)})}
\end{equation}
provides a stringent cross-check. Moreover, at LHCb, the $R_H$ ratios are measured as a double ratio between 
the rare and \jpsi resonant modes, to ensure any potential systematics in the computation of efficiencies cancel out.

\begin{figure}
	\centering
	{\includegraphics[width=0.4\linewidth]{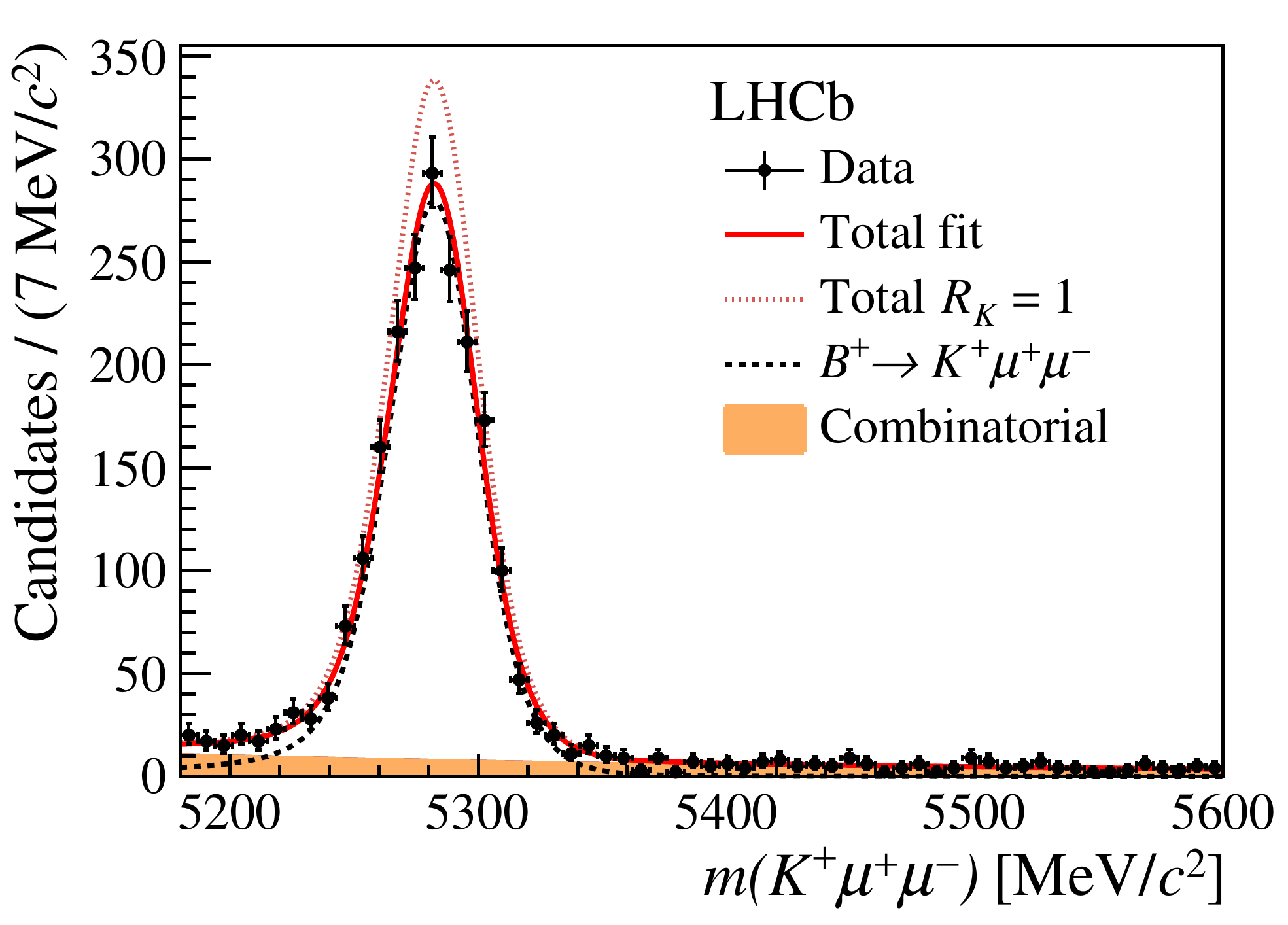}}
	{\includegraphics[width=0.4\linewidth]{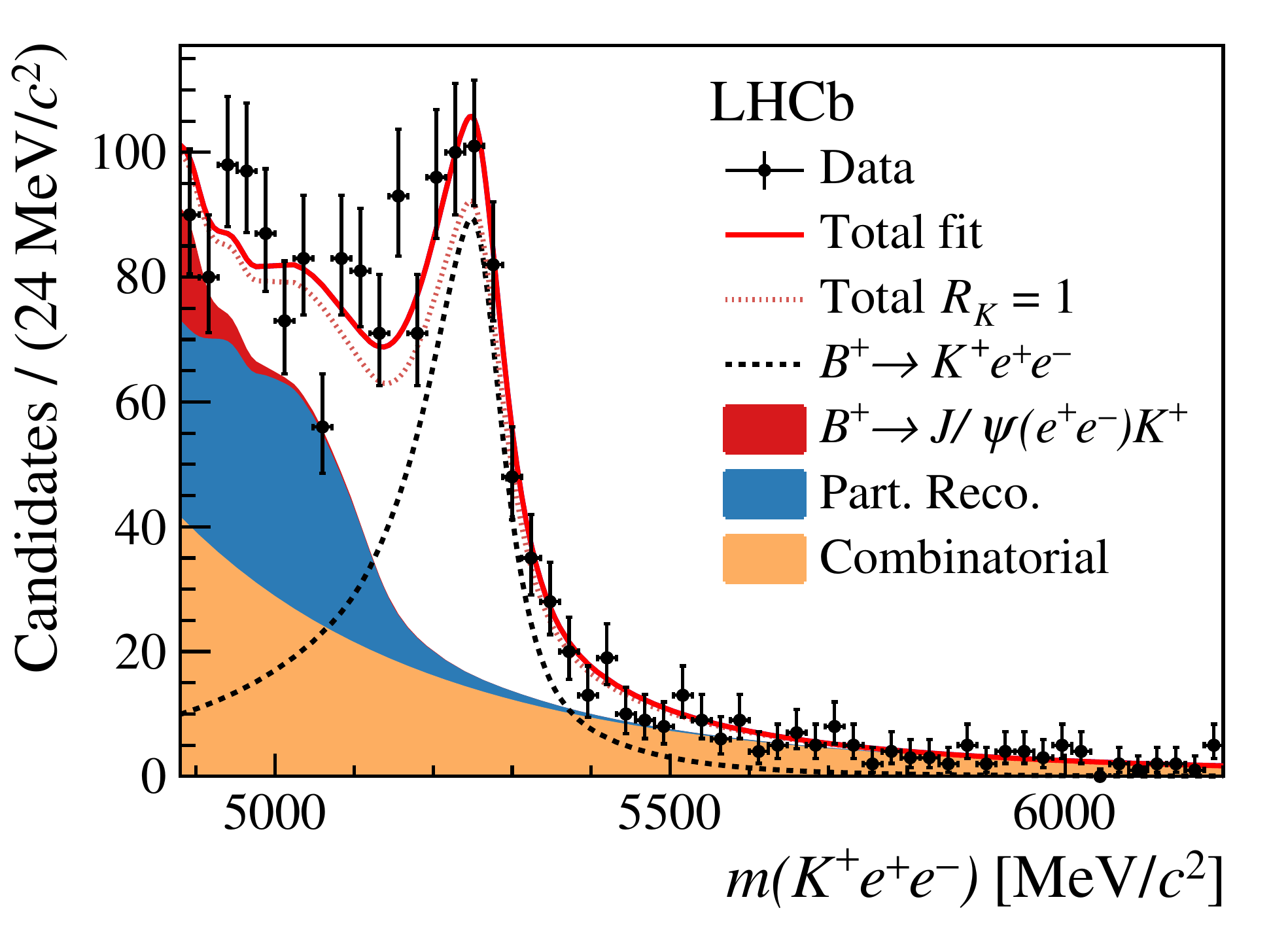}}
	\caption[]{Invariant mass of $\Kp\mumu$ (left) and $\Kp\epem$(right) data candidates (black points) with the result of the invariant 
		mass fit overlaid.}
	\label{fig:rk_lhcb_fits}
\end{figure}

An update of the measurement of the ratio $R_{K} = {\BR(\decay{\Bu}{\Kp \mumu})}/{\BR(\decay{\Bu}{\Kp \epem})}$ has been published 
by LHCb early this year~\cite{LHCb-PAPER-2019-009}. The analysis uses the dataset collected during 2011-2016, corresponding to an integrated luminosity of $5\invfb$, 
and measures $R_K$ in the region $1.1 < q^2 < 6.0\,\gevgevcccc$. A re-optimisation of the analysis of the 2011-2012 dataset~\cite{LHCb-PAPER-2014-024} 
is performed together with the inclusion of new data not previously studied. In total, a factor 2 larger statistics than in the previous 
analysis is obtained. The $r_{\jpsi}$ cross-check is found to be in good agreement with unity, $r_{\jpsi} = 1.014 \pm 0.035$. 
Moreover, this ratio is also measured in one and two-dimensional bins of decay kinematics and no significant trends are observed, 
validating the efficiency extraction from simulation and calibration samples. 
The invariant mass fits to $\Kp\mumu$ and $\Kp\epem$ candidates are shown in Fig.~\ref{fig:rk_lhcb_fits}. The broader signal shape for the electron 
mode requires the usage of a larger fit range and the inclusion of more backgrounds in the fit model. In particular, partially reconstructed 
events of the type \BdToKstee with $\decay{\Kstarz}{\Kp\pim}$, where the \pim is not reconstructed, and contamination from the \jpsi mode
due to large Bremsstrahlung loses are significant in this case. In total, around $1940$ and $760$ $\decay{\Bu}{\Kp \mumu}$ and 
$\decay{\Bu}{\Kp \epem}$ events are observed. Taking into account the reconstruction and selection efficiencies, the ratio of branching 
fractions between the two modes is measured to be 
\begin{equation}
R_{K} =  0.846 ^{+0.060 +0.016}_{-0.054-0.014},
\end{equation}
where the first uncertainty is statistical and the second systematic. This value is compatible with the previous LHCb result and deviates 
from the SM prediction of lepton universality at the level of $2.5\sigma$.

\begin{figure}
	\centering
	{\includegraphics[width=0.4\linewidth]{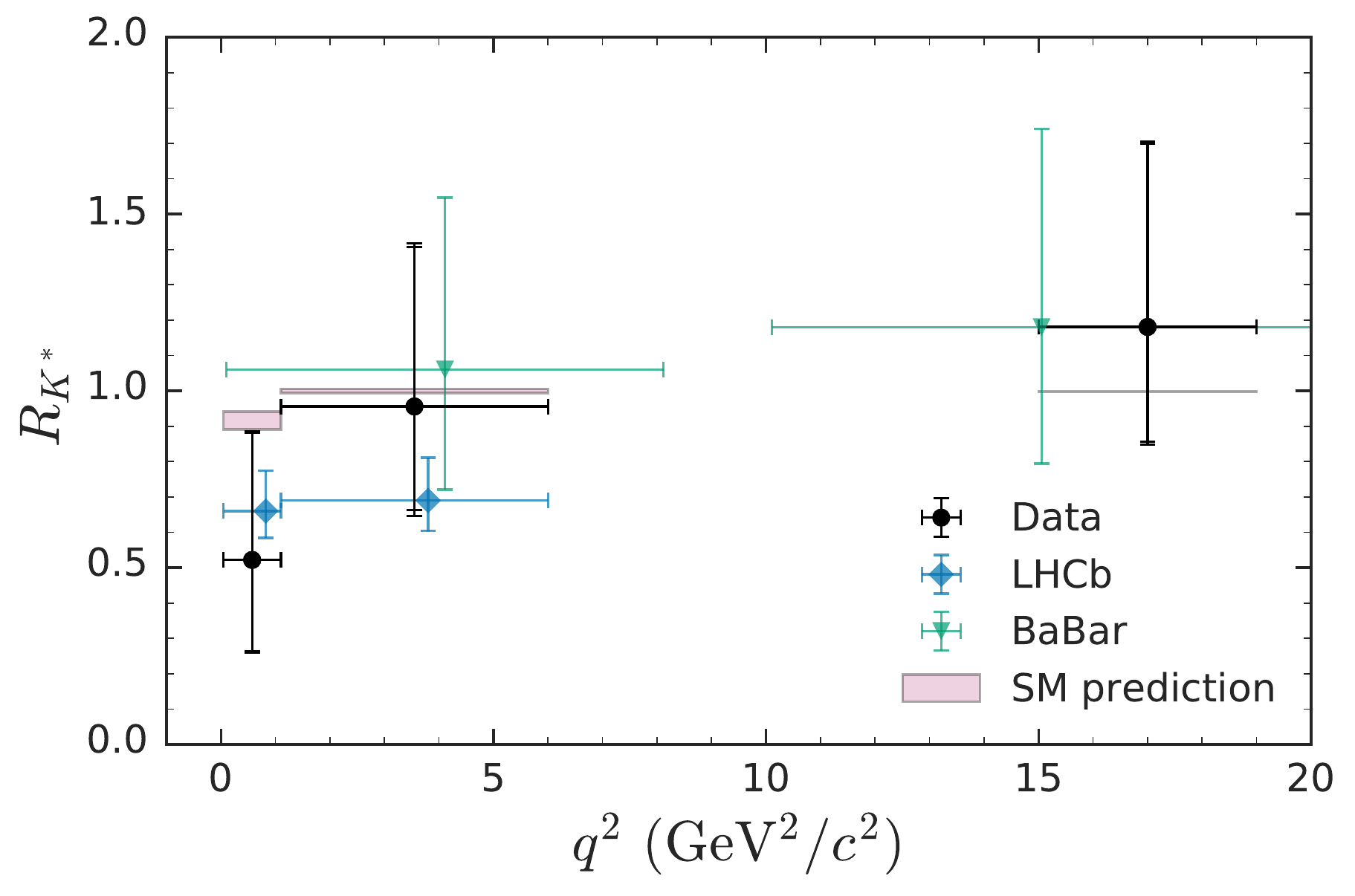}}
	\caption[]{Charge-averaged $R_{\Kstar}$ in bins of $q^2$ as measured by Belle.}
	\label{fig:belle_rkst}
\end{figure}

Belle has also recently published measurements of Lepton Universality with the ratios $R_{K}$ and 
$R_{\Kstar} = {\BR(\decay{\B}{\Kstar \mumu})}/{\BR(\decay{\Kstar}{\Kp \epem})}$~\cite{belle_rkst},
using both the charged and neutral modes and the full dataset. 
For $R_{\Kstar}$ the \Kstar meson is reconstructed in the final states $\Kp\pim$, $\KS\pim$ and $\Kp\piz$. 
The ratios involving neutral hadrons in the final state are measured for the first time in this analysis. 
As a cross-check, the branching fraction of the control mode $\decay{\B}{\Kstar\jpsi}$ is measured to be in agreement with the 
world average and the ratio $r_{\jpsi}$ is found to be in good agreement with unity, $r_{\jpsi} = 1.015 \pm0.025 \pm 0.038$, 
validating the efficiency determination. The signal yields are extracted from a fit to the beam-constrained invariant mass of $\Kstar\mumu$ and $\Kstar\epem$ candidates. 
The main backgrounds are combinatorial, from particle misidentification and from 
decays involving charmonium resonances. Around $140$ $\decay{\B}{\Kstar \mumu}$ and $100$ $\decay{\B}{\Kstar \epem}$
events are observed in total. The ratio of branching fractions $R_{\Kstar}$ is measured separately in the charged and neutral modes in various 
$q^2$ bins. All the values are found to be compatible with unity within uncertainties. The results are also compatible with the 
existing and more precise LHCb measurement~\cite{LHCb-PAPER-2017-013}.
The weighted average of the charged and neutral modes in the different $q^2$ regions is shown in Fig.~\ref{fig:belle_rkst}, together with 
the previous results from LHCb and BaBar and the SM predictions.

\begin{figure}
	\centering
	{\includegraphics[width=0.32\linewidth]{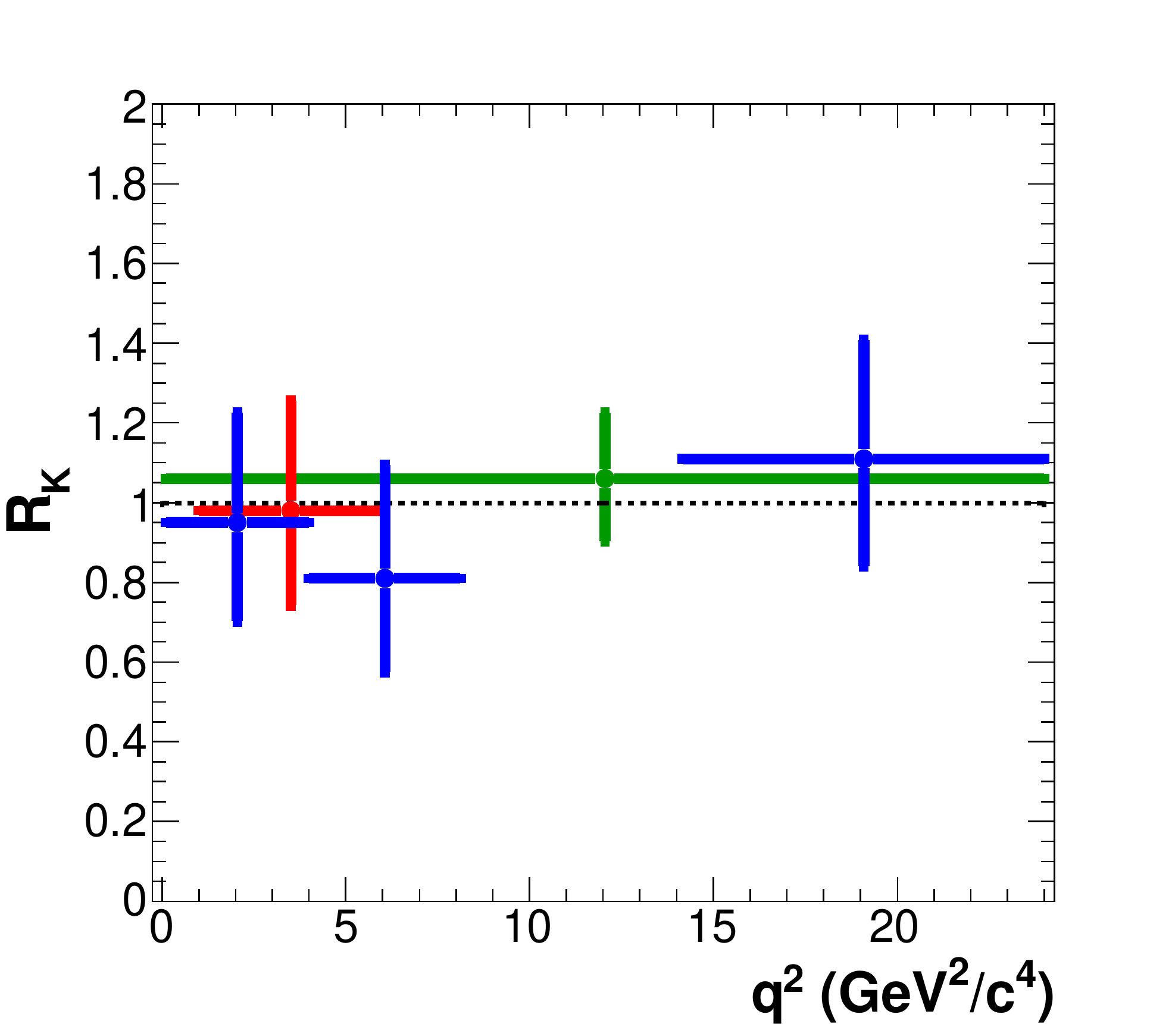}}
	{\includegraphics[width=0.32\linewidth]{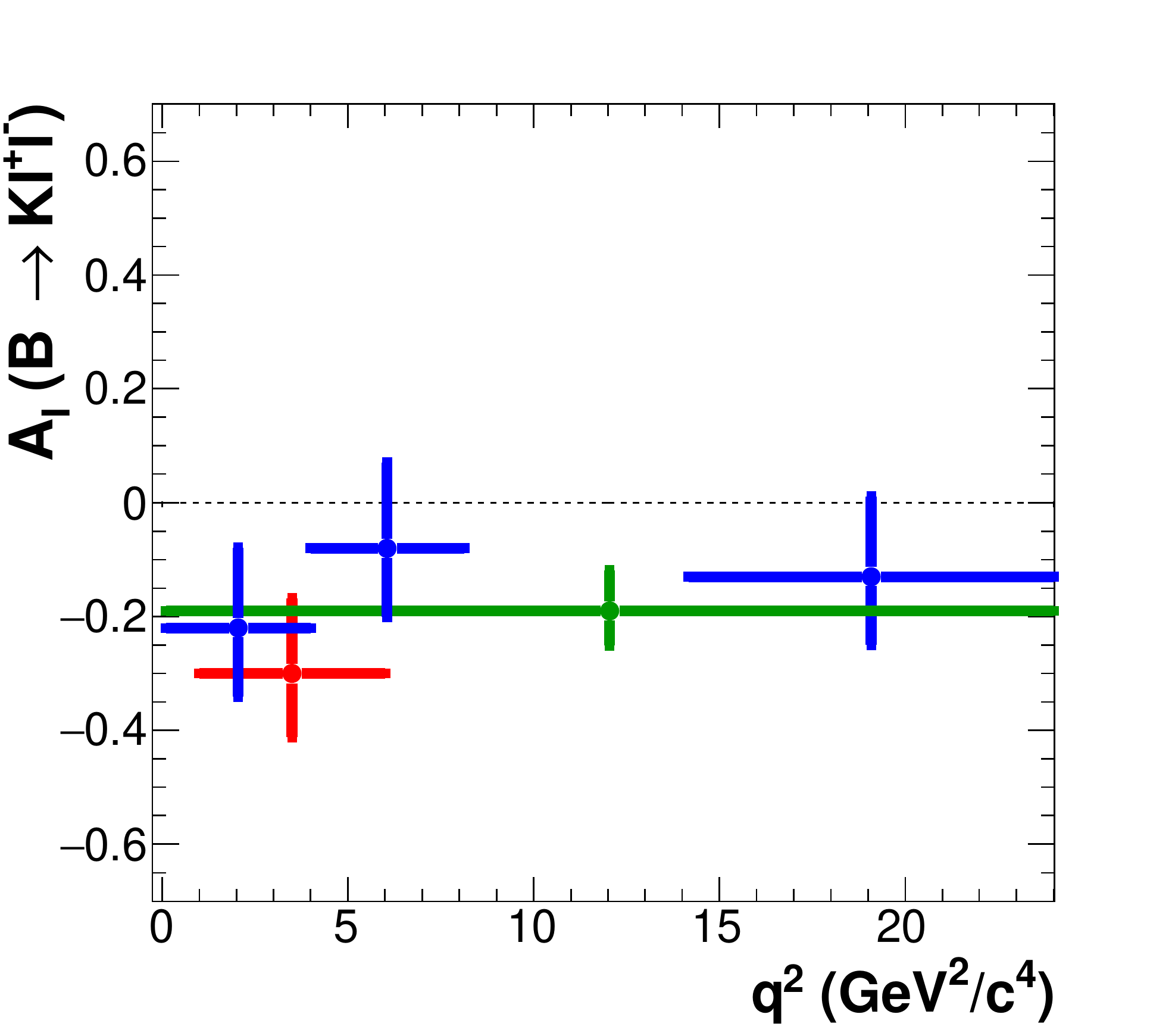}}
	\caption[]{Charge-averaged $R_{K}$ (left) and flavour-averaged $A_I$ (right) in bins of $q^2$ as measured by Belle.}
	\label{fig:belle_rk}
\end{figure}

The Belle $R_K$ measurement exploits a 3-dimensional fit of the beam-constrained mass, the energy difference between the \B candidate and
the beam and the output of the multivariate classifier used to separate signal and background. The measurement is also performed 
separately for the charged and neutral modes and in different $q^2$ regions. The charge average results are shown in Fig.~\ref{fig:belle_rk} (left)
and are found to be compatible with the SM and also with the LHCb measurement previously discussed. 
In the same analysis, Belle also measures the isospin asymmetry in 
\begin{equation}
	A_I = \frac{(\tau_{\Bu}/\tau_{\Bd}) \BR(\decay{\Bd}{\Kz\ell\ell}) -  \BR(\decay{\Bu}{\Kp\ell\ell})}
	{(\tau_{\Bu}/\tau_{\Bd}) \BR(\decay{\Bd}{\Kz\ell\ell}) +  \BR(\decay{\Bu}{\Kp\ell\ell})},
\end{equation}
where $(\tau_{\Bu}/\tau_{\Bd}) = 1.076$ is the lifetime ratio of \Bu to \Bd. This observable is also measured in bins of $q^2$. 
The results are found to be negative for almost all the $q^2$ bins in both electron and muon channels.
The combined results are shown in Fig.~\ref{fig:belle_rk} (right). 

\section{Lepton Flavour Violation searches}

Following the anomalies observed in \bsll transitions, several new physics (NP) models have been proposed to explain them. 
Specific examples can be found in Ref.~\cite{blanke}. Generally, models predicting lepton non-universality also imply lepton flavour 
violation, making the latter a sensitive probe to these NP scenarios.

The decay $\decay{\B _{(s)}}{\tau\mu}$ is forbidden in the SM but its branching fraction can be as large as $10^{-4}$ in NP models. 
For the \Bd mode a limit was set by BaBar at $\BR(\decay{\Bd}{\tau\mu}) < 2.2 \times 10^{-5}$ at $90\%$ CL~\cite{babar_taumu}, 
while no measurement is available for the \Bs mode. 
Using the full Run 1 data, corresponding to an integrated luminosity of $3\invfb$, LHCb has performed a search for this decay 
exploiting the hadronic tau decay $\decay{\taum}{\pim\pip\pim\neut}$ and kinematic constraints on the event~\cite{LHCb-PAPER-2019-016}.
No signal excess is observed in the region of interest so limits are set using the decay $\decay{\Bd}{\Dm (\to\Kp\pim\pim) \pip}$ as normalisation:
\begin{eqnarray}
 \BR(\decay{\Bd}{\tau\mu}) &<& 1.2 \times 10^{-5} \nonumber\\
 \BR(\decay{\Bs}{\tau\mu}) &<& 3.4 \times 10^{-5}
\end{eqnarray}
at $90\%$ CL, which are the most stringent limits to date. 

The decay $\decay{\Bu}{\Kp\muon\electron}$ is also forbidden in the SM and can reach the level of $10^{-8}$ in NP scenarios. 
Previous constraints on this decay from BaBar set limits on the branching fraction at $\BR(\decay{\Bu}{\Kp\mun\ep}) < 9.1\times 10^{-8}$ 
and $\BR(\decay{\Bu}{\Kp\mup\en}) < 13 \times 10^{-8}$ at $90\%$ CL~\cite{babar_kemu}. LHCb has very recently performed a search for this decay 
using the Run 1 data~\cite{LHCb-PAPER-2019-022}. The Bremsstrahlung recovery procedure described above is used to correct the electron momentum and a two-stage 
multivariate selection is employed to reduce the background from combinatorial and partially reconstructed \bquark-hadron decays.
No signal excess is observed and limits on the branching fraction are derived using the $\decay{\Bd}{\jpsi\Kp}$ decay
as normalisation:
\begin{eqnarray}
\BR(\decay{\Bu}{\Kp\mun\ep}) &<& 7.0 \times 10^{-9} \nonumber\\
\BR(\decay{\Bu}{\Kp\mup\en}) &<& 7.1 \times 10^{-9}
\end{eqnarray}
at $90\%$ CL, which are the most stringent limits to date, improving by more than an order of magnitude the previous constraints.

\section{Summary}

Rare decays of b-hadrons provide high sensitivity to NP effects. Several deviations with respect to the SM predictions have been 
observed in recent years, leading to significant tensions in global fit analyses.
The latest results from \lhcb and Belle on radiative, penguin semileptonic and lepton flavour violation decays are presented.
These involve improvements on the constraints on $C'_7$, complementary angular distribution measurements and further tests 
of lepton flavour universality in \bsll transitions and new or largely improved constraints on lepton flavour violating decays. 
The tensions in \bsll processes with respect to the SM predictions remain after these results, motivating further updates with larger 
datasets and the exploration of new modes. LHCb has already collected data corresponding to $9\invfb$, a good part of which remains 
to be analysed, and both Belle II and LHCb will collect much larger datasets in the coming years, which will allow to clearly disentangle 
the presence and nature of new physics if the current observed anomalies are indeed originated by physics beyond the
SM~\cite{belleII_book,lhcb_upgrade2}.

\end{document}

%% file: symbols.tex
\usepackage{xspace} 
\usepackage{upgreek} 

\def\PB      {\ensuremath{B}\xspace}                 
\def\PD      {\ensuremath{D}\xspace}                 
\def\PJ      {\ensuremath{{J}}\xspace}          
\def\PK      {\ensuremath{{K}}\xspace}          
       
\def\Pb      {\ensuremath{{b}}\xspace}

\def\Pe      {\ensuremath{{e}}\xspace}
\def\Pnu         {\ensuremath{\upnu}\xspace}
\def\Pp      {\ensuremath{{p}}\xspace}
\def\Ps      {\ensuremath{{s}}\xspace}

\def\Pgamma      {\ensuremath{\gamma}\xspace}
\def\Pmu         {\ensuremath{\upmu}\xspace} 
                 
\def\Ppi         {\ensuremath{\pi}\xspace}                 
\def\Ppsi        {\ensuremath{\psi}\xspace}                 
\def\Ptau        {\ensuremath{\tau}\xspace}

\mathchardef\PXi="7104
\mathchardef\PLambda="7103
\mathchardef\POmega="710A

\def\squark    {{\ensuremath{\Ps}}\xspace}
\def\bquark    {{\ensuremath{\Pb}}\xspace}

\def\pion   {{\ensuremath{\Ppi}}\xspace}
\def\piz    {{\ensuremath{\pion^0}}\xspace}

\def\kaon   {{\ensuremath{\PK}}\xspace}
\def\Kz       {{\ensuremath{\kaon^0}}\xspace}
\def\Kstar  {{\ensuremath{\kaon^{*}}}\xspace}
\def\Kstarz {{\ensuremath{\kaon^{*0}}}\xspace}

\def\Kp       {{\ensuremath{\kaon^+}}\xspace}

\def\KS       {{\ensuremath{\kaon^0_{\rm\scriptscriptstyle S}}}\xspace}

\def\D       {{\ensuremath{\PD}}\xspace}

\def\Dm      {{\ensuremath{\D^-}}\xspace}

\def\B       {{\ensuremath{\PB}}\xspace}
\def\Bd      {{\ensuremath{\B^0}}\xspace}
\def\Bs      {{\ensuremath{\B^0_\squark}}\xspace}
\def\Bu      {{\ensuremath{\B^+}}\xspace}

\def\proton      {{\ensuremath{\Pp}}\xspace}

\def\Lz          {{\ensuremath{\PLambda}}\xspace}

\def\Lb      {{\ensuremath{\Lz^0_\bquark}}\xspace}

\def\pion   {{\ensuremath{\Ppi}}\xspace}
\def\pip    {{\ensuremath{\pion^+}}\xspace}
\def\pim    {{\ensuremath{\pion^-}}\xspace}

\def\electron   {{\ensuremath{\Pe}}\xspace}
\def\en         {{\ensuremath{\Pe^-}}\xspace}   
\def\ep         {{\ensuremath{\Pe^+}}\xspace}
 
\def\epem       {{\ensuremath{\Pe^+\Pe^-}}\xspace}

\def\muon       {{\ensuremath{\Pmu}}\xspace}
\def\mup        {{\ensuremath{\Pmu^+}}\xspace}
\def\mun        {{\ensuremath{\Pmu^-}}\xspace} 
\def\mumu       {{\ensuremath{\Pmu^+\Pmu^-}}\xspace}

\def\taum       {{\ensuremath{\Ptau^-}}\xspace}

\def\ellm       {{\ensuremath{\ell^-}}\xspace}
\def\ellp       {{\ensuremath{\ell^+}}\xspace}

\def\neu        {{\ensuremath{\Pnu}}\xspace}

\def\neut       {{\ensuremath{\neu_\tau}}\xspace}

\def\g      {{\ensuremath{\Pgamma}}\xspace}

\def\jpsi     {{\ensuremath{{\PJ\mskip -3mu/\mskip -2mu\Ppsi\mskip 2mu}}}\xspace}
\def\psitwos  {{\ensuremath{\Ppsi{(2S)}}}\xspace}

\newcommand{\decay}[2]{\ensuremath{#1\!\to #2}\xspace}         

\def\bsll         {\decay{\bquark}{\squark\ellp\ellm}}

\def\bsg          {\decay{\bquark}{\squark\gamma}}

\def\BdToKstmm   {\decay{\Bd}{\Kstarz\mup\mun}}

\def\LbToLzmm    {\decay{\Lb}{\Lz\mup\mun}}

\def\BdToKstee   {\decay{\Bd}{\Kstarz\epem}}

\def\BdKstGam     {\decay{\Bd}{\Kstarz \g}}

\def\bsphig      {\decay{\Bs}{\phi\g}}
\def\lblg           {\decay{\Lb}{\Lz\g}}
\def\LbLzGam  {\decay{\Lb}{\Lz \g}}
\def\LbLzEta  {\decay{\Lb}{\Lz \eta}}

\def\LbtoLz {\decay{\Lb}{\Lz}}

\def\BF {{\ensuremath{\mathcal{B}}}\xspace}
\def\BR {\BF}

\def\BRlblg {\ensuremath{\BF(\lblg)}\xspace}
\def\BRLbLzGam {\ensuremath{\BF(\LbLzGam)}\xspace}

\def\lhcb {\mbox{LHCb}\xspace}

\def\invfb   {\ensuremath{\mbox{\,fb}^{-1}}\xspace}
\newcommand{\tev}{\ensuremath{\mathrm{\,Te\kern -0.1em V}}\xspace}
\newcommand{\mev}{\ensuremath{\mathrm{\,Me\kern -0.1em V}}\xspace}
\newcommand{\mevc}{\ensuremath{{\mathrm{\,Me\kern -0.1em V\!/}c}}\xspace}
\newcommand{\mevcc}{\ensuremath{{\mathrm{\,Me\kern -0.1em V\!/}c^2}}\xspace}
\newcommand{\gev}{\ensuremath{\mathrm{\,Ge\kern -0.1em V}}\xspace}
\newcommand{\gevc}{\ensuremath{{\mathrm{\,Ge\kern -0.1em V\!/}c}}\xspace}
\newcommand{\gevcc}{\ensuremath{{\mathrm{\,Ge\kern -0.1em V\!/}c^2}}\xspace}
\newcommand{\gevgevcc}{\ensuremath{{\mathrm{\,Ge\kern -0.1em V^2\!/}c^2}}\xspace}
\newcommand{\gevgevcccc}{\ensuremath{{\mathrm{\,Ge\kern -0.1em V^2\!/}c^4}}\xspace}

